\documentclass[12pt]{article}
\usepackage{graphicx}
\usepackage{url}
\usepackage{lineno}

\newcommand\pubnumber{}
\newcommand\pubdate{\today}

\def\milano{Istituto Nazionale di Fisica Nucleare\\
Sezione di Milano, I-20131 Milano, ITALY}
\def\behalf{\footnote{on behalf of the \babar collaboration.}}

\def\Title#1{\begin{center} {\Large #1 } \end{center}}
\def\Author#1{\begin{center}{ \sc #1} \end{center}}
\def\Address#1{\begin{center}{ \it #1} \end{center}}

\newcommand\pubblock{\rightline{\begin{tabular}{l} \pubnumber\\
         \pubdate  \end{tabular}}}
\newenvironment{Abstract}{\begin{quotation}  }{\end{quotation}}
\newenvironment{Presented}{\begin{quotation} \begin{center} 
             PRESENTED AT\end{center}\bigskip 
      \begin{center}\begin{large}}{\end{large}\end{center} \end{quotation}}
\def\Acknowledgements{\bigskip  \bigskip \begin{center} \begin{large}
             \bf ACKNOWLEDGEMENTS \end{large}\end{center}}
\input econfmacros.tex

\begin{document}
\begin{titlepage}
\pubblock

\vfill
\Title{Recent results for \Dz-\Dzb mixing and \CP violation, and HFAG averages}
\vfill
\Author{ Nicola Neri\behalf}
\Address{\milano}
\vfill
\begin{Abstract}
I present here the HFAG averages for the parameters that regulate flavor mixing and \CP violation
 in the neutral $D$ meson system. I also discuss recent results from the $B$ factories
 for the measurements of the mixing parameter \yCP\ and the \CP violation parameter \deltaY\ (\AGamma) 
in the lifetime ratio analysis of the transitions to the \CP-even eigenstates $\Dz\to K^+K^-,\pi^+\pi^-$, 
 relative to the transitions to the \CP-mixed state $\Dz \to K^-\pi^+$.
\end{Abstract}
\vfill
\begin{Presented}
The 5th International Workshop on Charm Physics\\ (Charm 2012)\\
Honolulu, Hawai'i (USA), May 14-17, 2012
\end{Presented}
\vfill
\end{titlepage}
\def\thefootnote{\fnsymbol{footnote}}
\setcounter{footnote}{0}
%

\section{Introduction}
Several measurements have reported evidence for flavor mixing in the neutral $D$ meson 
system~\cite{Aubert:2007wf,Aubert:2007en,Aubert:2008zh,Aubert:2009ai,Staric:2007dt,Aaltonen:2007ac}. 
The results are in agreement with the Standard Model (SM) predictions, which unfortunately are affected 
 by large theoretical uncertainties~\cite{Xing:1996pn,Bianco:2003vb,Burdman:2003rs}. 
However, the increasing precision of \Dz-\Dzb mixing measurements 
 helps constrain new physics models~\cite{Golowich:2007ka,Golowich:2009ii}. 
\CP violation in $D$ meson decays, though notoriously difficult to calculate precisely,
  is expected to be very small in the SM, at the level of $10^{-3}$ or 
less~\cite{Bianco:2003vb,Buccella:1994nf,Grossman:2006jg}. 
Relatively large \CP asymmetries, at the percent level, might be a signature of new physics effects.
Recent results from the LHCb experiment~\cite{Aaij:2011in} reported evidence for direct \CP violation 
 measuring the difference of \CP asymmetries in singly-Cabibbo suppressed  $\Dz \to \pi^+\pi^-$ 
\footnote{The use of charge conjugate reactions is implied throughout unless stated otherwise.} 
and $\Dz \to K^+K^-$ decays, with a statistical significance of $3.5\sigma$. 
The observed asymmetries are 
 marginally compatible with the SM but not conclusive for establishing new 
physics~\cite{Grossman:2012eb,Franco:2012ck,Isidori:2011qw}.
These intriguing results renew interest in studying mixing 
and \CP violation in the \DzDzb meson system and in general in the charm physics sector.

The Heavy Flavor Averaging Group (HFAG)~\cite{HFAG:webpage} provides averages for heavy flavor quantities. 
It is divided into several sub-groups, each of which focuses on a different set of heavy flavor measurements. 
 The Charm Physics sub-group~\cite{HFAG_charm:webpage} studies the following topics: \Dz-\Dzb mixing, 
\CP violation, spectroscopy of charm mesons and baryons, semileptonic decays, decay constants, 
hadronic branching fraction measurements and rare $D$ decay modes.
In particular, it provides averages for the the \Dz-\Dzb mixing and \CPV parameters by combining 
 measurements from different experiments in a $\chi^2$-based fit. 
The results of the fit are expressed in term of physics parameters that can be directly 
 compared to the theoretical predictions. 
\section{HFAG notations for flavor mixing and \boldmath{\CPV} }
Flavor mixing occurs when the Hamiltonian eigenstates (or physics eigenstates)  
$D_1$, $D_2$ differ from the flavor eigenstates \Dz, \Dzb. 
A neutral $D$ meson produced at time $t=0$ in a definite flavor state \Dz, 
will then evolve and oscillate into a state of opposite flavor \Dzb, after a certain time.
If we describe the time evolution of the flavor eigenstates in terms of
 a $2\times 2$ effective Hamiltonian, 
$\boldmath{H} = \boldmath{M} - \frac{i}{2} \boldmath{\Gamma}$, then, assuming \CPT is conserved, the mass eigenstates can be 
 expressed in terms of the flavor eigenstates by,
\beqa
|D_1 \rangle = p|\Dz\rangle + q |\Dzb\rangle \\
|D_2 \rangle = p|\Dz\rangle - q |\Dzb\rangle \nonumber
\eeqan
with the normalization $|q|^2+|p|^2=1$ and 
\beq
\frac{q}{p} = \sqrt{\frac{M_{12}^*-i/2\Gamma_{12}^*}{M_{12}-i/2\Gamma_{12}}}.
\eeqn
Assuming a phase convention such that $CP |\Dz \rangle = -|\Dzb \rangle$ and 
$CP |\Dzb \rangle = -|\Dz\rangle$ then, if \CP is conserved, we have that $q=p=1/\sqrt2$ and the 
 mass eigenstates coincide with the \CP eigenstates: $|D_1\rangle = |D_{CP_-}\rangle$ (\CP-odd) and $|D_2\rangle = |D_{CP_+}\rangle$ (\CP-even).

The mixing parameters can be expressed in terms of the difference of masses ($m_{1,2}$) and widths ($\Gamma_{1,2}$) of the Hamiltonian eigenstates,
\beq
x =  \frac{m_2 - m_1}{\Gamma}, \hspace{2cm}
y =  \frac{\Gamma_2 - \Gamma_1}{2\Gamma}, 
\eeqn
where $\Gamma = (\Gamma_1+\Gamma_2)/2$.

\CP violation can be of three types:
\begin{enumerate}
\item{\it \CPV in decay or direct \CPV}: this occurs when the decay amplitudes
for \CP conjugate processes are different in modulus. 
If $\langle f|H|\Dz\rangle = A_f$, $\langle \bar{f}|H|\Dzb\rangle = \bar{A}_{\bar{f}}$ are the
 \Dz and \Dzb decay amplitudes into the final states $f$ and \CP conjugate $\bar{f}$, then
\beq 
\left|\frac{\bar{A}_{\bar{f}}}{A_f}\right| \ne 1 \Longrightarrow \CPV.
\eeqn
\item{\it \CPV in mixing or indirect \CPV}: it occurs when the Hamiltonian eigenstates 
do not coincide with the \CP eigenstates. That is 
\beq
\left| \frac{q}{p} \right| \ne 1 \Longrightarrow \CPV.
\eeqn
\item{\it \CPV in the interference of mixing and decay}: 
for neutral $D$ mesons there is a third possibility to observe \CP violation even when \CP 
is conserved in mixing and also in decay. In this case, \CP violation arises when, 
in a process with final state $f$ that can be reached by  neutral $D$ mesons of both flavors 
(\ie~\Dz and \Dzb),  there is a relative weak phase
\footnote{The \CP-violating phase is also indicated as a weak phase since it originates from the weak interaction
 in the SM.} difference between the mixing and the decay amplitudes.
The quantity of interest that is independent of phase conventions, and physically meaningful, is
\beq \label{eq:lambda}
\lambda_f \equiv {q\over p}{\bar A_{f}\over A_{f}} \equiv \left|{q\over p}\right|\left|{\bar A_{f}\over A_{f}} \right| e^{i(\delta_f+\phi_f)},
\eeqn
where $\delta_f$ and $\phi_f$ are the \CP-conserving and \CP-violating phases respectively.
If \CP is conserved in mixing and in decay, the signature of \CP violation in the interference of mixing and decay is thus
\beq
\phi_f \ne 0 \Longrightarrow \CPV.
\eeqn
For \CP eigenstates, \CPV in either mixing or decay is indicated by
\beq
|\lambda_f| \ne 1,
\eeqn
while \CPV in the interference of mixing and decay corresponds to
\beq
\textrm{Im}(\lambda_f) \ne 1.
\eeqn
Note that if there is no weak phase in the decay amplitudes then $\arg{(q/p)} = \phi$ and
 it is independent of the final state $f$. 
\end{enumerate}

\section{Experimental observables and parameters of the theory}
The most precise constraints on the mixing parameters \x, \y and the \CPV parameters \absqop, 
$\phi$ are obtained in time-dependent analyses of \D decays. 

Consider a final state $f$ that can be reached by both \Dz and \Dzb decays. If a \Dz is produced at time $t=0$, 
it can reach the final state $f$ by mixing to $\Dzb$ followed by the decay $\Dzb \to f$, 
or directly through the decay $\Dz \to f$. The interference between the mixing and decay amplitudes modifies the time-dependence 
 with respect to the pure exponential as follows:
\begin{eqnarray}
\frac{|\langle f | H | \Dz(t)\rangle|^2}{\frac12e^{-\Gamma t }}  &=& 
 \left( \absAf^2+\mabsqop^2\absAbf^2 \right) \cosh(y \Gamma  t) + 
   2\textrm{Re} \left( \mqop \Af^*\Abf \right)  \sinh(y \Gamma t)  \nonumber  \\
& + & \left( \absAf^2 - \mabsqop^2 \absAbf^2 \right) \cos(x \Gamma  t) -
 2\textrm{Im} \left( \mqop \Af^*\Abf  \right) \sin (x \Gamma t)   \\
\frac{|\langle f | H | \Dzb(t)\rangle|^2}{\frac12e^{-\Gamma t }}  &=& 
 \left( \absAbf^2+\mabspoq^2\absAf^2 \right) \cosh(y \Gamma  t) + 
   2\textrm{Re} \left( \mpoq \Af\Abf^* \right)  \sinh(y \Gamma t)  \nonumber  \\
& + & \left( \absAbf^2 - \mabspoq^2 \absAf^2 \right) \cos(x \Gamma  t) -
 2\textrm{Im} \left( \mpoq \Af\Abf^*  \right) \sin (x \Gamma t).   
\end{eqnarray}
HFAG combines 38 observables measured in time-dependent and time-integrated analyses from the following experiments: 
\babar, Belle, CDF, CLEO, CLEOc, E791, FOCUS, and LHCb. When allowing for \CPV there are 10 underlying parameters that 
 are extracted from the $\chi^2$ fit: \x, \y, \absqop, $\phi$, $\delta$, $\delta_{K\pi\pi}$, $R_D$, $A_D$, $A_\pi$, $A_K$.
The parameters $\delta$, $\delta_{K\pi\pi}$ are relative strong phases, $R_D$ is the ratio $\Gamma(\Dz\to K^+\pi^-)/\Gamma(\Dzb\to K^-\pi^+)$ 
 and $A_D$, $A_\pi$, $A_K$ are the direct \CP-violation asymmetries 
for the $\Dz \to K^+\pi^-$, $\Dz \to \pi^+\pi^-$ and $\Dz \to K^+K^-$ modes.
The relationships between these parameters and the measured observables are given below.
\begin{enumerate}
\item  {\it Semileptonic decays}: search for mixing by reconstructing the ``wrong-sign'' (WS) decay chain, $D^{*+} \to \Dz \pi^+$, $\Dz \to \Dzb$, 
 $\Dzb \to K^{(*)+}e^-\bar{\nu}_e$. In contrast to hadronic decays, the WS charge combinations can occur only through mixing.
The measurement of $R_M$ is related to the mixing parameters as follows
\beq
R_M = \frac12(x^2+y^2),
\eeqn
and can be obtained directly as the ratio of WS to right-sign (RS) signal events.
The RS events correspond to the non-mixed process.
\item {\it Decays to \CP eigenstates}: measure the mixing parameter \yCP\ and the \CPV parameter $A_\Gamma$ with a lifetime ratio analysis 
of the transitions to the \CP eigenstates and the transitions to the CP-mixed state $\Dz \to K^-\pi^+$,
\beqa
2 \yCP = \left(\mabsqop + \mabspoq \right) y \cos\phi - \left( \mabsqop -\mabspoq \right) x \sin \phi \\ \nonumber
2 A_\Gamma = \left(\mabsqop - \mabspoq \right) y \cos\phi - \left( \mabsqop +\mabspoq \right) x \sin \phi .
\eeqan
The parameter $A_\Gamma$ is the decay-rate asymmetry for the \CP eigenstates. If \CP is conserved $\yCP = y$ and $A_\Gamma = 0$. 

Time-integrated \CP-violating asymmetries in \CP-even eigenstates, \eg\ in $\Dz \to K^+K^-$ and $\Dz \to \pi^+\pi^-$, provide constraints on the 
 mixing and \CPV parameters according to the relations,
\beqa
\frac{\Gamma(\Dz \to K^+K^-) + \Gamma(\Dzb \to K^-K^+) }{\Gamma(\Dz \to K^-K^+) + \Gamma(\Dzb \to K^+K^-)} =  A_K + \frac{\langle t \rangle}{\tau_D} A_{\CP}^{\scriptsize \textrm{indirect}}\\ \nonumber
\frac{\Gamma(\Dz \to \pi^+\pi^-) - \Gamma(\Dzb \to \pi^-\pi^+) }{\Gamma(\Dz \to \pi^-\pi^+) + \Gamma(\Dzb \to \pi^+\pi^-)} = A_\pi + \frac{\langle t \rangle}{\tau_D} A_{\CP}^{\scriptsize \textrm{indirect}}, \nonumber
\eeqan
where
\beq
2 A_{\CP}^{\scriptsize \textrm{indirect}} = \left(\mabsqop + \mabspoq \right) x \sin\phi - \left( \mabsqop -\mabspoq \right) y \cos \phi, 
\eeqn
$\langle t \rangle$ is the average reconstructed \Dz proper time and $\tau_D$ is the nominal \Dz lifetime.
\item {\it Three-body $\Dz \to K_S^0\pi^+\pi^-$and $\Dz \to K_S^0K^+K^-$ decays}: measure directly the mixing 
and \CPV parameters \x, \y, \absqop, and $\phi$ with  a time-dependent Dalitz plot analysis. 
%
\item {\it Wrong-sign decays to hadronic non-CP eigenstates}: measure the parameters $x'^{\pm}$ and $y'^{\pm}$
  and $R_D$ and $A_D$ in a  time-dependent analysis of the WS events selected through the decay chain 
$D^{*+}\to\Dz \pi^+$, $\Dz \to K^+\pi^-$. 
The parameters are defined as follows:
\beqa
x'^{\pm} = \left( \frac{1\pm A_M}{1\mp A_M}\right)^{1/4} (x'\cos\phi\pm y' \sin\phi) \\  \nonumber
y'^{\pm} = \left( \frac{1\pm A_M}{1\mp A_M}\right)^{1/4} (y'\cos\phi\mp x' \sin\phi) \\ \nonumber
\frac{\Gamma(\Dz \to K^+\pi^-) + \Gamma(\Dzb \to K^-\pi^+) }{\Gamma(\Dz \to K^-\pi^+) + \Gamma(\Dzb \to K^+\pi^-)} = R_D \\ \nonumber
\frac{\Gamma(\Dz \to K^+\pi^-) - \Gamma(\Dzb \to K^-\pi^+) }{\Gamma(\Dz \to K^-\pi^+) + \Gamma(\Dzb \to K^+\pi^-)} = A_D  \nonumber
\eeqan
where
\beqa
\left( \begin{array}{c} x' \\ y' \end{array} \right) & = &
\left( \begin{array}{cc} 
\cos\delta & \sin\delta \\ \nonumber
-\sin\delta & \cos\delta \end{array} \right)
\left( \begin{array}{c} x \\ y \end{array} \right) \\ \nonumber
A_M  =  \frac{\absqop^2 - \abspoq^2}{\absqop^2 + \abspoq^2}, &&\hspace{2cm}  
\delta = \arg\left( \frac{A(\Dz \to K^+ \pi^-)}{A(\Dzb \to K^+ \pi^-)}\right). 
\eeqan
The superscript in $x'^{\pm}$ and $y'^{\pm}$ identifies the flavor of the \D sample, 
\ie\ \Dz ($^{+}$) and \Dzb ($^{-}$).

\par The mixing parameters $x''$ and $y''$,
\beqa
\left( \begin{array}{c} x'' \\ y'' \end{array} \right) & = &
\left( \begin{array}{cc} 
\cos\delta_{K\pi\pi} & \sin\delta_{K\pi\pi} \\ \nonumber
-\sin\delta_{K\pi\pi} & \cos\delta_{K\pi\pi} \end{array} \right)
\left( \begin{array}{c} x \\ y \end{array} \right)  \nonumber
\eeqan
may be measured by means of a time-dependent Dalitz plot analysis of the three-body WS events $\Dz \to K^+\pi^-\pi^0$. 
The strong phase that rotates the mixing  parameters \x and \y is defined here as  
$\delta_{K\pi\pi} = \arg\left( \frac{A(\Dz \to K^+ \rho^-)}{A(\Dzb \to K^+ \rho^-)}\right)$.
\item {\it $\psi(3770) \to \DzDzb$ reaction}: 
exploiting the quantum correlations of the \DzDzb production 
at threshold, it is possible to constrain the 
 physics parameters by measuring $x^2$, $y$, $R_D$, $2 \sqrt{R_D} \cos\delta$, $2 \sqrt{R_D} \cos\delta$.
\end{enumerate}

\section{HFAG results for \boldmath{\Dz-\Dzb} mixing and \boldmath{\CPV}}
HFAG uses a global fit to determine world averages of mixing parameters,
\CPV parameters, and strong phases. 
The fit uses 38 observables taken from measurements of $\Dz \to K^+\ell^-\nu$, 
$\Dz \to K^+K^-$, $\Dz \to \pi^+\pi^-$, $\Dz \to K^+\pi^-$,  $\Dz \to K^+\pi^-\pi^0$, 
$\Dz \to K_S^0\pi^+\pi^-$, $\Dz \to K_S^0K^+K^-$ decays and from double-tagged branching fractions measured 
at the $\psi(3770)$ resonance. Correlations among observables are accounted for by using covariance matrices 
provided by the experimental collaborations. Errors are assumed to be Gaussian, and systematic errors among different
experiments are assumed uncorrelated unless specific correlations have been identified. An independent 
log-likelihood fit which accounts for non-Gaussian errors has been used as a control check, 
and equivalent results  have been obtained.

\begin{figure}
\begin{center}
\includegraphics[scale=0.5]{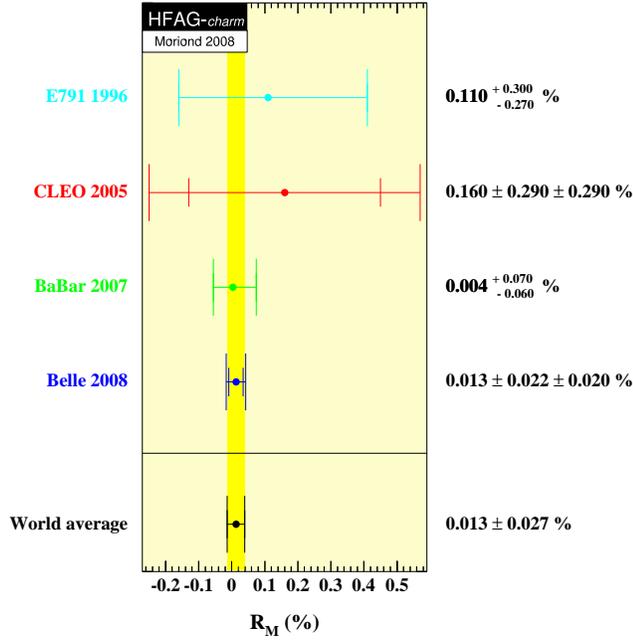}
\end{center}
\caption{\label{fig:rm_semi}
World average value of $R_M$ from Ref.~\cite{HFAG_charm:webpage},
as calculated from $D^0\to K^+\ell^-\nu$ 
measurements~\cite{Aitala:1996vz,Cawlfield:2005ze,Aubert:2007aa,Bitenc:2008bk}. }
\end{figure}

\begin{figure}
\begin{center}
\includegraphics[scale=0.45]{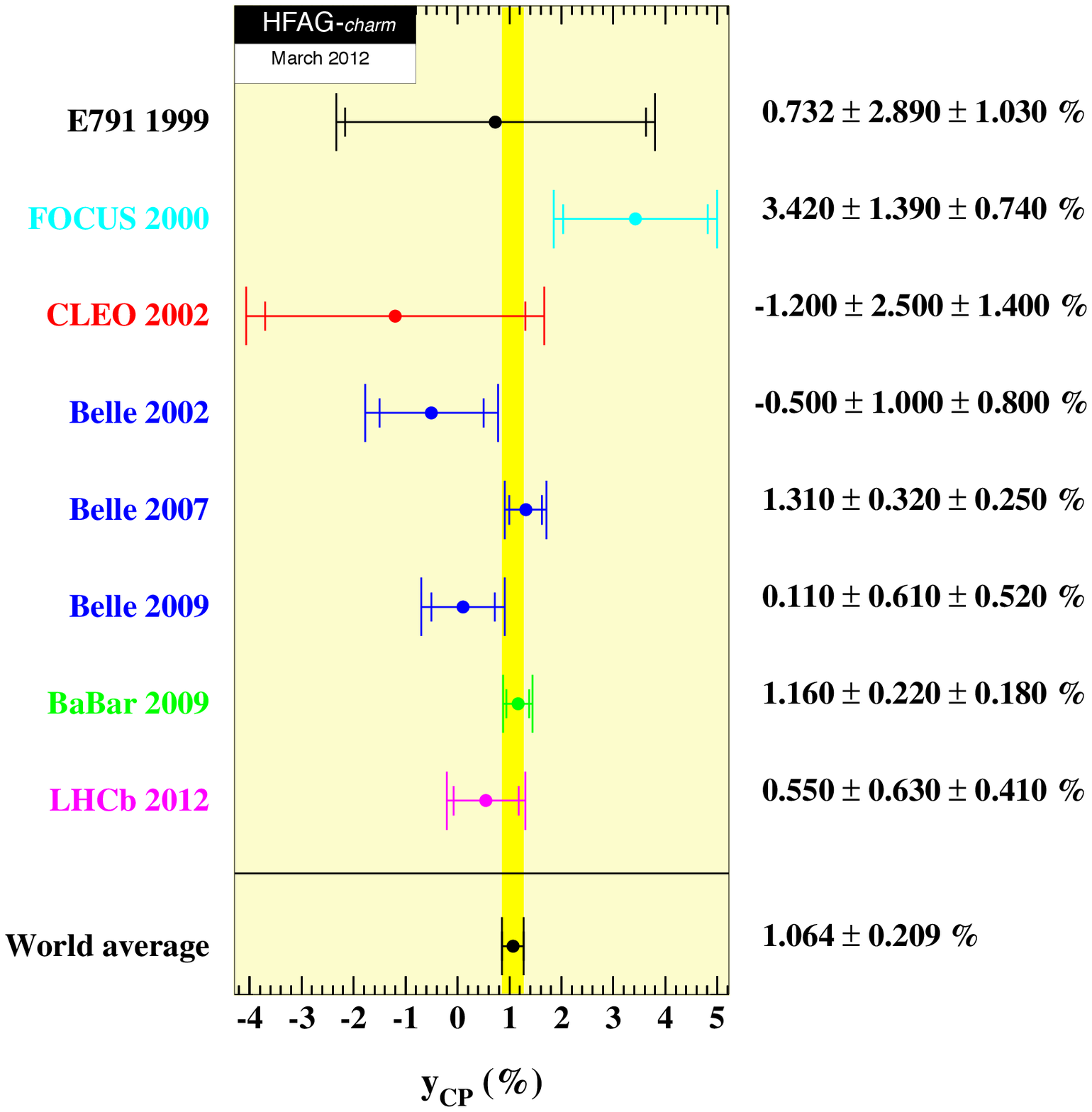}
\includegraphics[scale=0.45]{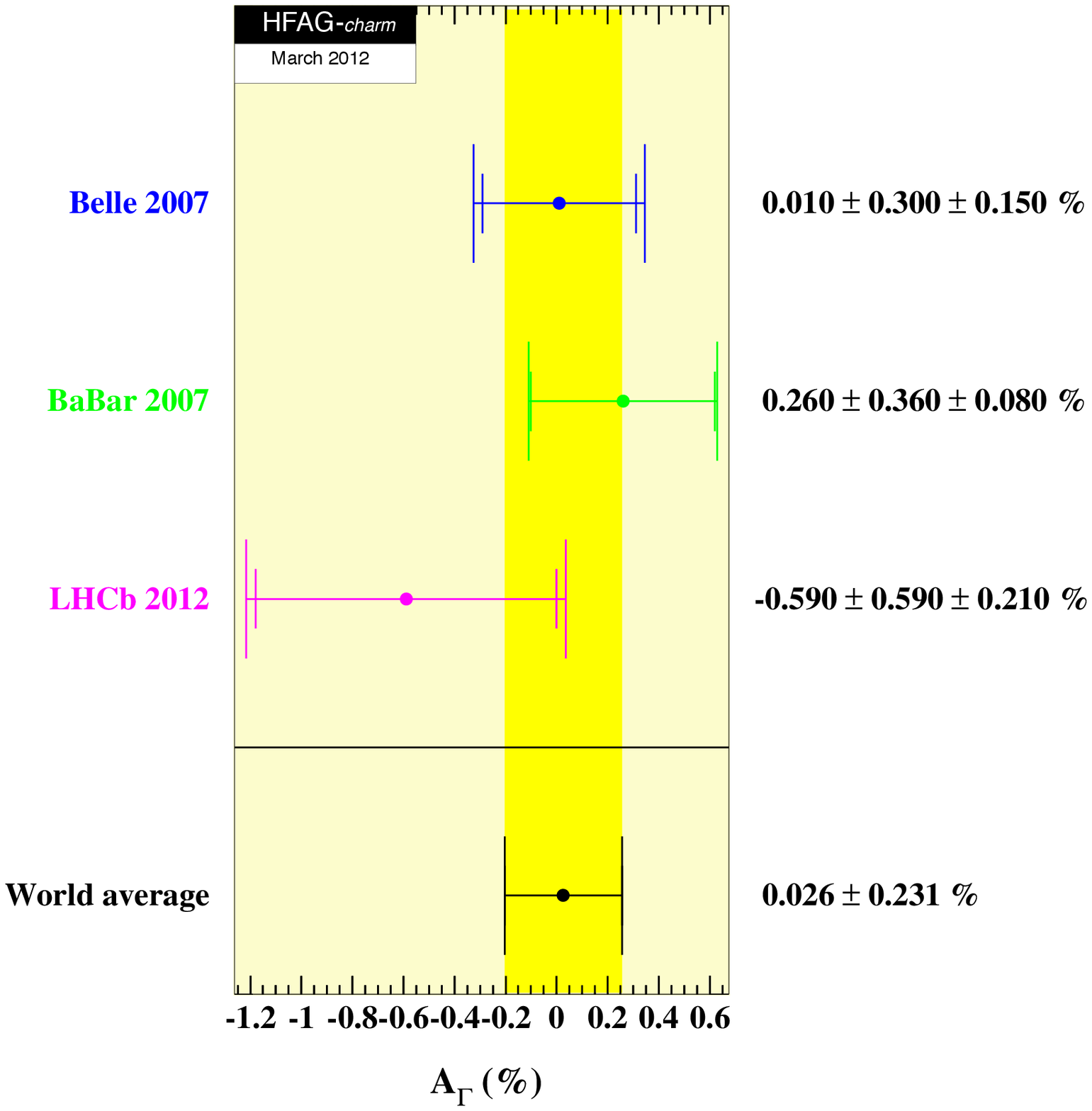}
\end{center}
\caption{\label{fig:ycp_Agamma}
World average value of $y^{}_{CP}$ and $A_\Gamma$ from Ref.~\cite{HFAG_charm:webpage}, 
as calculated from \dkkpp.}
\end{figure}

Averages for the parameters $R_M$, \yCP and \AGamma are calculated and then provided as input to the global fit.
The observable $R_M$ is calculated from $\Dz \to K^+\ell^-\nu$ 
decays~\cite{Aitala:1996vz,Cawlfield:2005ze,Aubert:2007aa,Bitenc:2008bk},
 and the average value is 
$R_M = (0.013\pm0.027)\%$. The inputs used for this average are plotted in Fig.~\ref{fig:rm_semi}.
The observable \yCP is calculated from \dkkpp\ and from $\Dz \to K_S^0K^+K^-$ decays while 
 \AGamma is calculated from \dkkpp\ decays only. 
The inputs used for these averages are plotted in Fig.~\ref{fig:ycp_Agamma}.
The average values are $\yCP = (1.064\pm0.209)\%$  and $\AGamma = (0.026\pm 0.231)\%$.

The \dkpi\ observables used are from Belle~\cite{Zhang:2006dp}, 
\babar~\cite{Aubert:2007wf}, and CDF~\cite{Aaltonen:2007ac};
earlier measurements have much less precision and are not used.
The observables from \dkspp\ decays for no-\CPV\ are from 
Belle~\cite{Abe:2007rd} and \babar~\cite{delAmoSanchez:2010xz}, 
but for the \CPV-allowed case only Belle measurements~\cite{Abe:2007rd} 
are available. The $D^0\ra K^+\pi^-\pi^0$ results are from 
\babar~\cite{Aubert:2008zh}, and the $\psi(3770)\ra\DzDzb$
results are from CLEOc~\cite{Sun:2010zz}. 
Time-integrated measurements of \CP-violating asymmetries ($A_{\CP}$) in the $\Dz \to K^+K^-$ and $\Dz \to \pi^+\pi^-$ modes
are from \babar~\cite{Aubert:2007if} and Belle~\cite{Staric:2008rx}. The difference of the \CP-violating asymmetries defined 
as $\Delta A_{\CP} = A_{\CP}(K^+K^-)-A_{\CP}(\pi^+\pi^-)$ are from LHCb~\cite{Aaij:2011in} and CDF~\cite{CDF:2012qw}. 
These measurements reported evidence for direct \CPV with a statistical significance 
of $3.5\sigma$ and  $2.7\sigma$, respectively.\\
\par
Three types of fit are produced by HFAG with different \CPV assumptions:
\begin{enumerate}

\item {\it No \CPV}: in this fit it is assumed that \CP is conserved and the \CPV parameters $A_D$, $A_K$, $A_\pi$, \absqop-1 and $\phi$ are fixed to zero.
\item {\it No direct \CPV}: the \CPV parameters $A_D$, $A_K$, $A_\pi$ are fixed to zero. 
In this case the relation~\cite{Ciuchini:2007cw,Kagan:2009gb} 
$\tan\phi = (1-|q/p|^2)/(1+|q/p|^2)\times (x/y)$ is satisfied, and this reduces 
the four independent parameters (\x, \y, \absqop, $\phi$) to three.
The independent parameters used in the fit are $x^{}_{12}= 2|M^{}_{12}|/\Gamma$,  $y^{}_{12}= \Gamma^{}_{12}/\Gamma$, and 
$\phi^{}_{12}= {\rm Arg}(M^{}_{12}/\Gamma^{}_{12})$, where $M^{}_{12}$ and $\Gamma^{}_{12}$ are the off-diagonal
elements of the $\Dz-\Dzb$ mass and decay matrices, respectively.
\item {\it \CPV-allowed}: where all the parameters for mixing and \CPV are floated in the fit.
\end{enumerate}

\begin{figure}
\begin{center}
\includegraphics[scale=0.5]{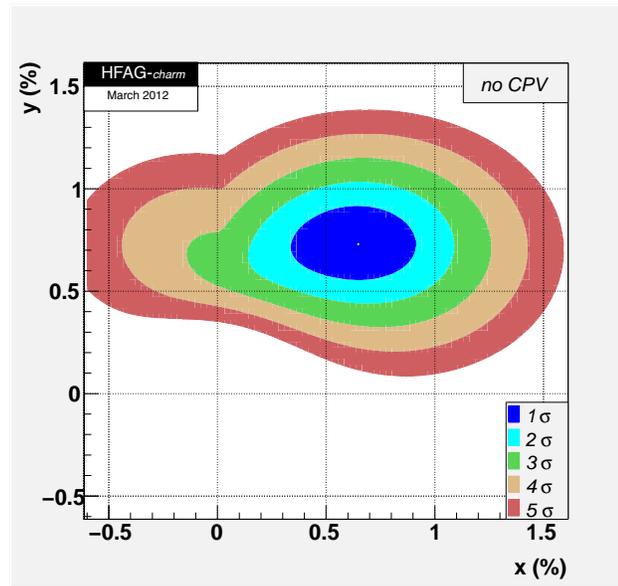}
\end{center}
\caption{\label{fig:contours_ncpv}
Two-dimensional contours for the mixing parameters $(x,y)$, for no \CPV. }
\end{figure}

\begin{figure}
\begin{center}
\vbox{
\includegraphics[scale=0.45]{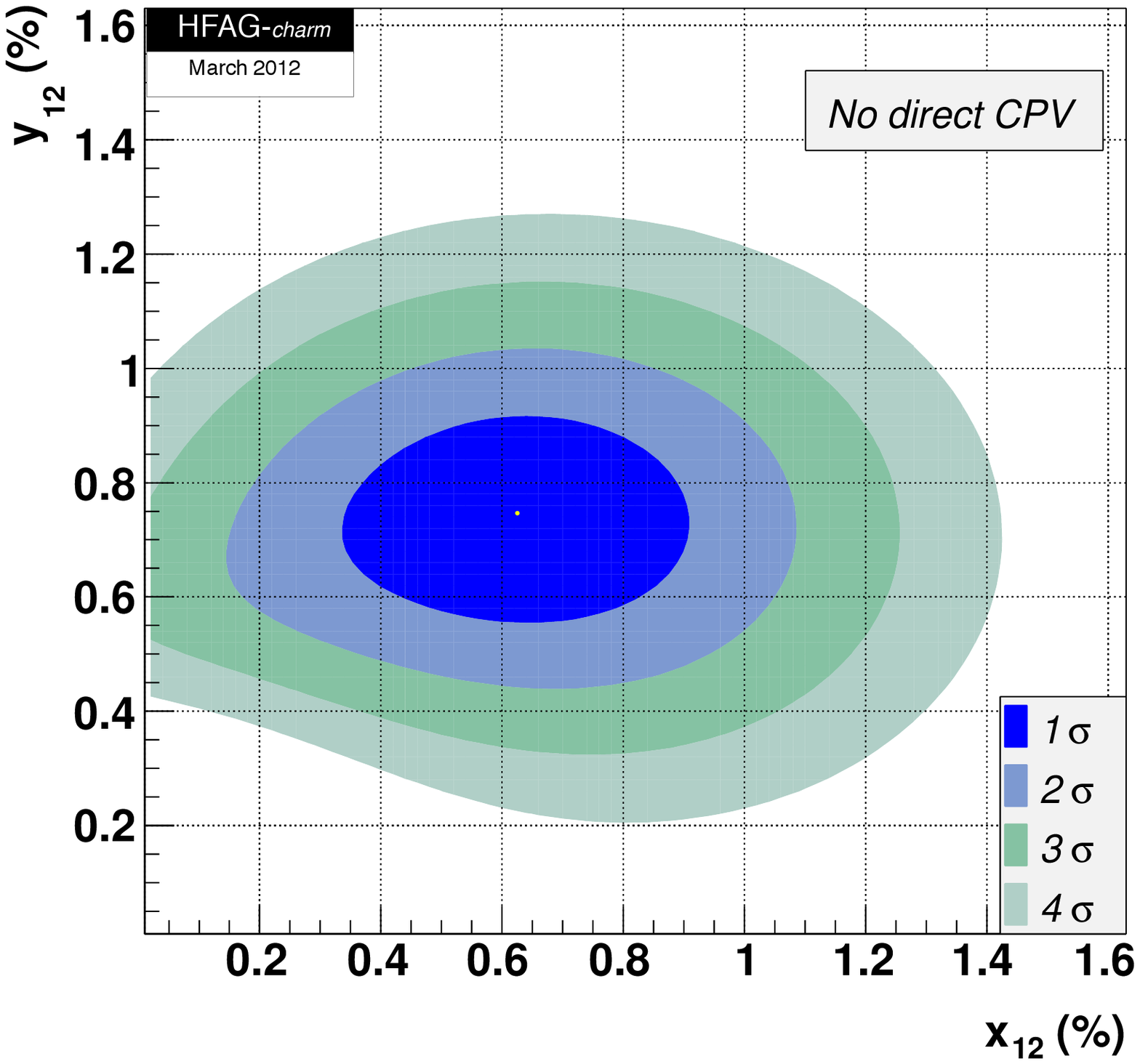}
\includegraphics[scale=0.45]{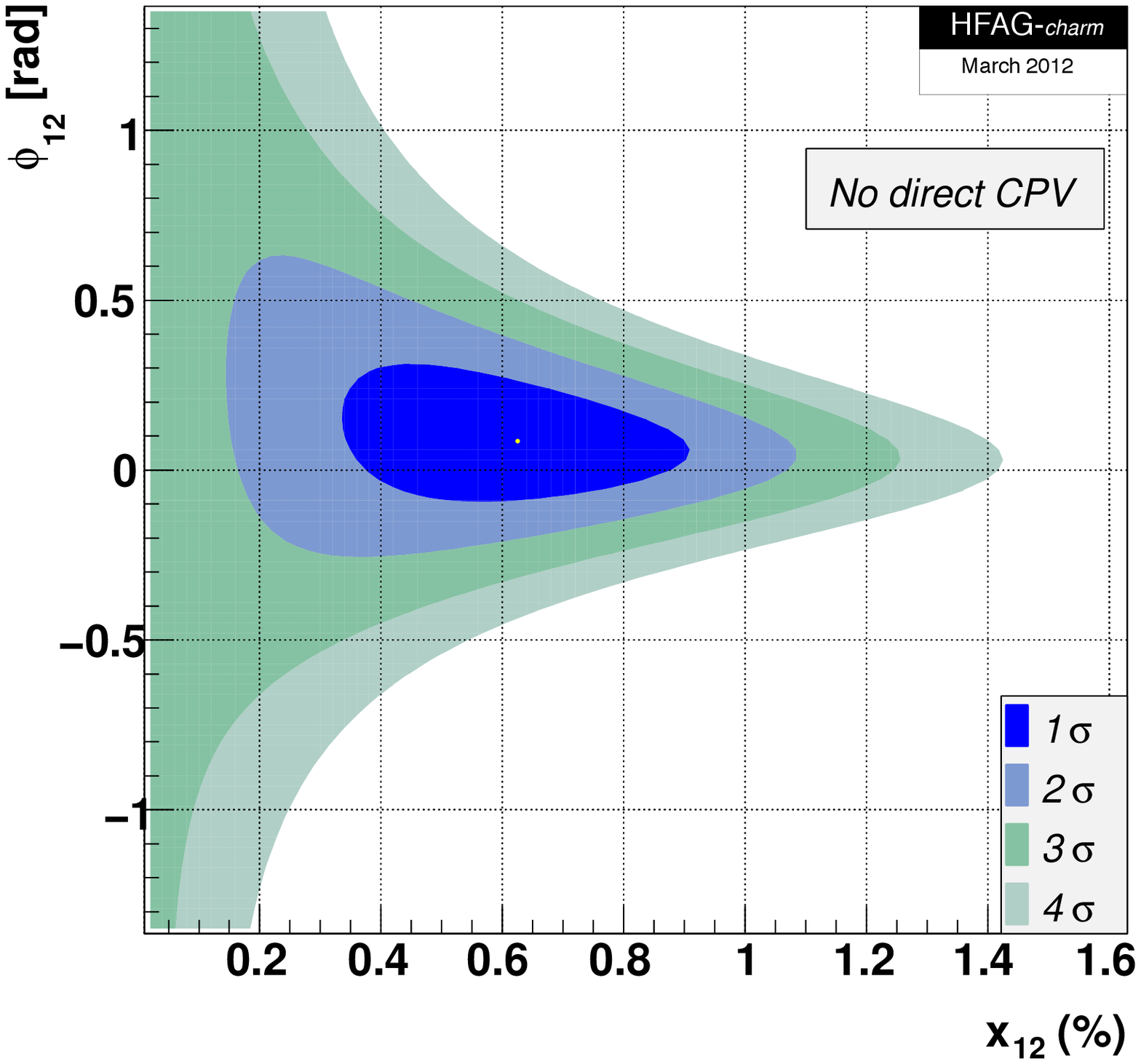}
\vskip0.30in
\includegraphics[scale=0.45]{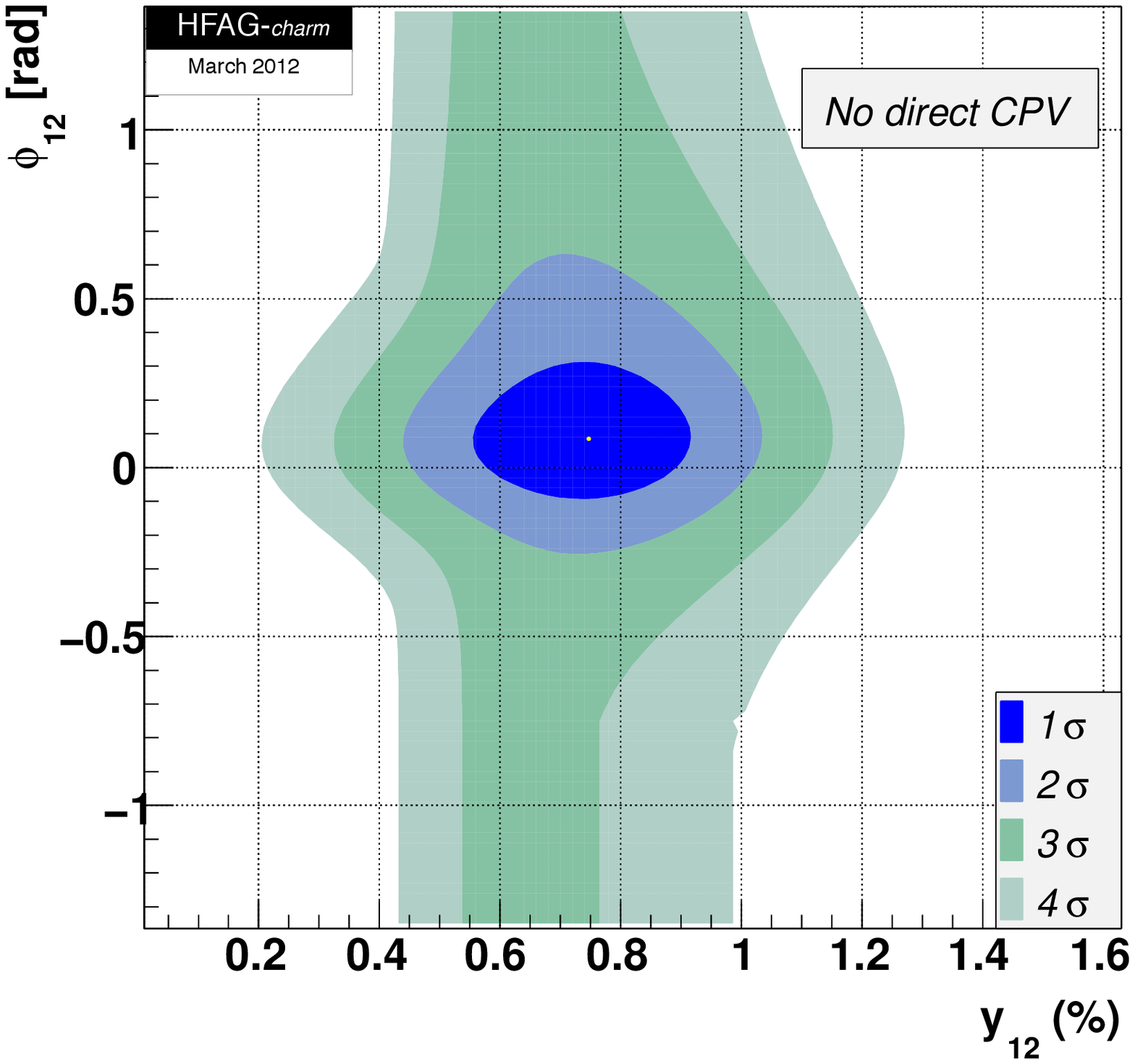}
}
\end{center}
\vskip-0.10in
\caption{\label{fig:contours_ndcpv}
Two-dimensional contours for theory parameters 
$(x^{}_{12},y^{}_{12})$ (top left), 
$(x^{}_{12},\phi^{}_{12})$ (top right), and 
$(y^{}_{12},\phi^{}_{12})$ (bottom), 
for no direct \CPV.}
\end{figure}

\begin{figure}
\begin{center}
\includegraphics[scale=0.45]{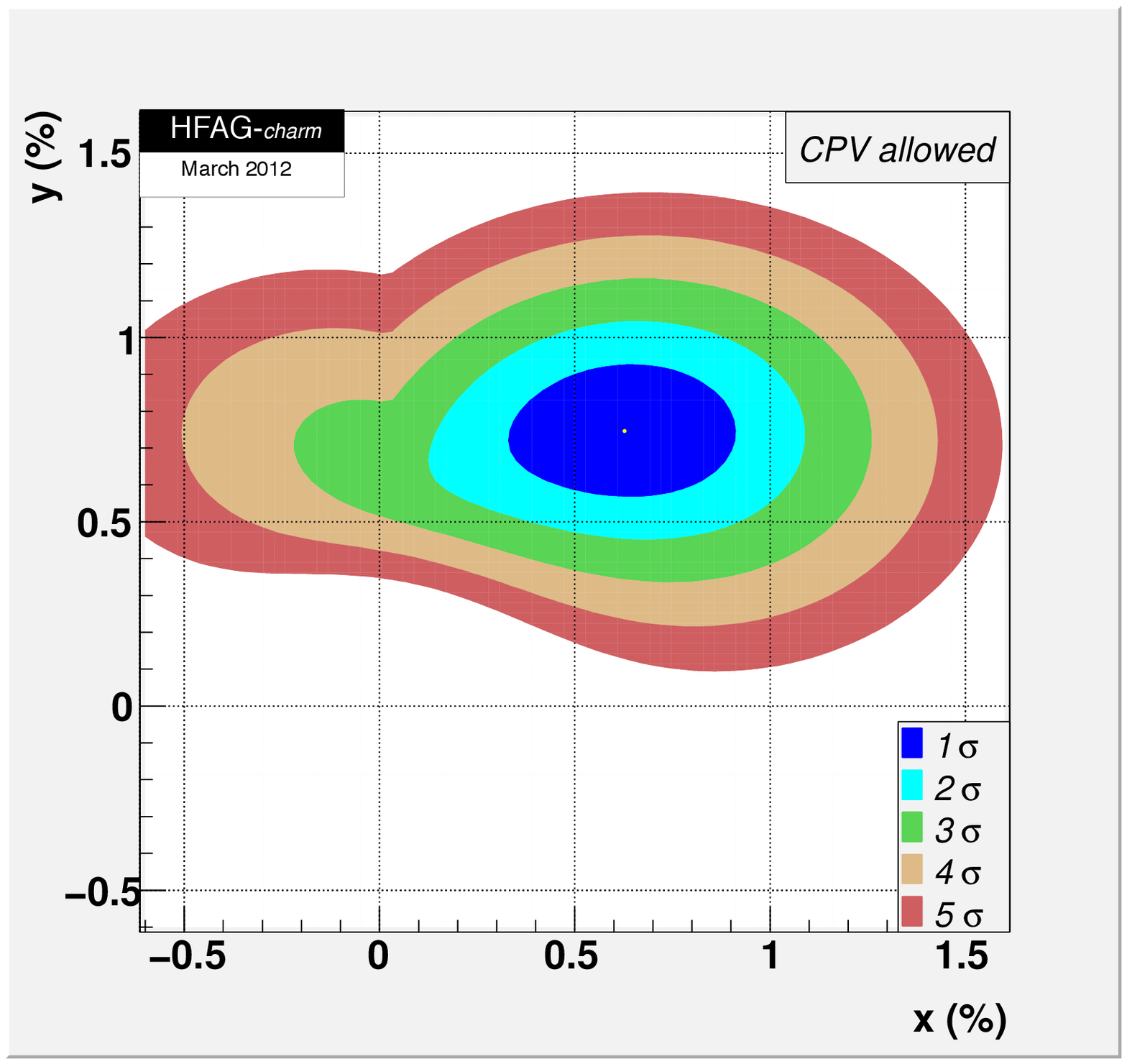}
\includegraphics[scale=0.45]{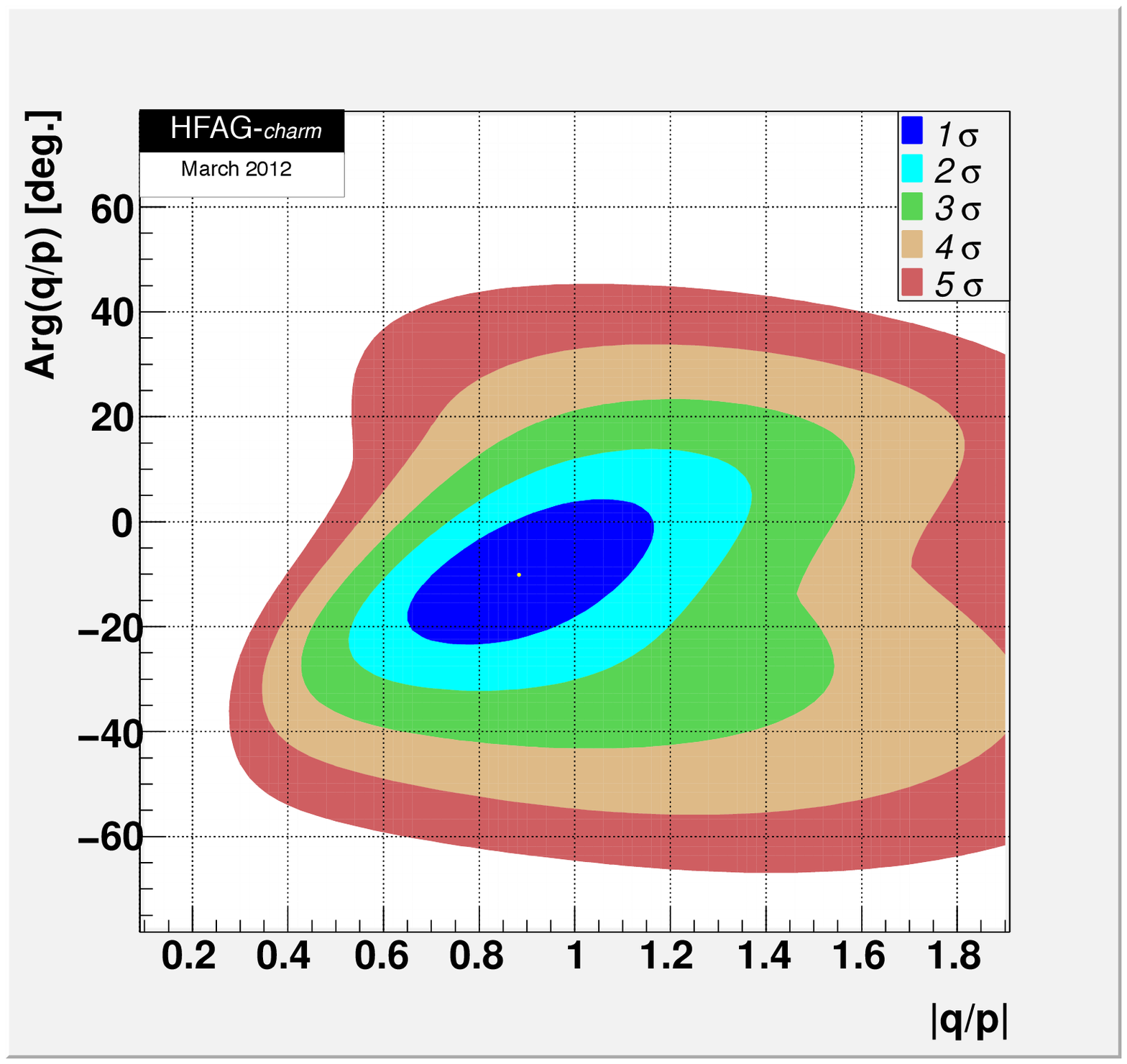}
\end{center}
\vskip-0.10in
\caption{\label{fig:contours_cpv}
Two-dimensional contours for the parameters $(x,y)$ (left) 
and $(|q/p|,\phi)$ (right), allowing for \CPV.}
\end{figure}

All fit results are listed in 
Table~\ref{tab:results}. 
The total $\chi^2$  is 35.6 for $37-10=27$ degrees of freedom; this 
corresponds to a confidence level of~12.4\%.
The resulting $1\sigma$-$5\sigma$ contours  are shown 
in Fig.~\ref{fig:contours_ncpv} for the \CP-conserving case, 
in Fig.~\ref{fig:contours_ndcpv} for the no-direct-\CPV\ case, 
and in Fig.~\ref{fig:contours_cpv} for the \CPV-allowed 
case. For the \CPV-allowed fit, the no-mixing point $(\x, \y) = (0, 0)$ is excluded at 
 a confidence level of $1.28\times 10^{-24}$ corresponding to a statistical significance of $10.2\sigma$.
The parameter \x differs from zero by $2.7\sigma$, and \y differs from zero by $6.0\sigma$.
In the $(|q/p|,\phi)$ plot, the point $(1,0)$ 
is within the $1\sigma$ contour; thus the data are consistent with \CP\ conservation in mixing and 
 in the interference between mixing and decay.

\begin{table}
\renewcommand{\arraystretch}{1.4}
\begin{center}
\caption{\label{tab:results}
Results of the global fit for different assumptions concerning~\CPV.}
\vspace*{6pt}
\footnotesize
\begin{tabular}{c|cccc}
\hline
\textbf{Parameter} & \textbf{\boldmath No \CPV} & \textbf{\boldmath No direct \CPV} 
& \textbf{\boldmath \CPV-allowed} & \textbf{\boldmath \CPV-allowed 95\% C.L.}  \\
\hline
$\begin{array}{c}
x\ (\%) \\ 
y\ (\%) \\ 
\delta\ (^\circ) \\ 
R^{}_D\ (\%) \\ 
A^{}_D\ (\%) \\ 
|q/p| \\ 
\phi\ (^\circ) \\
\delta^{}_{K\pi\pi}\ (^\circ)  \\
A^{}_{\pi} \\
A^{}_K \\
x^{}_{12}\ (\%) \\ 
y^{}_{12}\ (\%) \\ 
\phi^{}_{12} (^\circ)
\end{array}$ & 
$\begin{array}{c}
0.65\,^{+0.18}_{-0.19} \\
0.73\,\pm 0.12 \\
21.0\,^{+9.8}_{-11.0} \\
0.3307\,\pm 0.0080 \\
0.0 \\
1.0 \\
0.0 \\
17.8\,^{+21.7}_{-22.8} \\
0.0 \\
0.0 \\
- \\
- \\
- 
\end{array}$ &
$\begin{array}{c}
0.62\,\pm 0.19 \\
0.75\,\pm 0.12 \\
22.2\,^{+9.9}_{-11.2} \\
0.3305\,\pm 0.0080 \\
0.0 \\
1.04\,^{+0.07}_{-0.06} \\ 
-2.02\,^{+2.67}_{-2.74} \\ 
19.4\,^{+21.8}_{-22.9} \\
0.0 \\
0.0 \\
0.62\,\pm 0.19 \\
0.75\,\pm 0.12 \\
4.9\,^{+7.7}_{-6.5} 
\end{array}$ &
$\begin{array}{c}
0.63\,^{+0.19}_{-0.20}  \\
0.75\,\pm 0.12 \\
22.1\,^{+9.7}_{-11.1} \\
0.3311\,\pm 0.0081 \\
-1.7\,\pm 2.4 \\
0.88\,^{+0.18}_{-0.16} \\
-10.1\,^{+9.5}_{-8.9} \\ 
19.3\,^{+21.8}_{-22.9} \\
0.36\,\pm 0.25 \\
-0.31\,\pm 0.24 \\
- \\
- \\
- 
\end{array}$ &
$\begin{array}{c}
\left[0.24 ,\, 0.99\right] \\
\left[0.51 ,\, 0.98\right] \\
\left[-2.6 ,\, 40.6\right] \\
\left[0.315 ,\, 0.347\right] \\
\left[-6.4 ,\, 3.0\right] \\
\left[0.59 ,\, 1.26\right] \\
\left[-27.4 ,\, 8.7\right] \\
\left[-26.3 ,\, 61.8\right] \\ 
\left[-0.13 ,\, 0.86\right] \\
\left[-0.78 ,\, 0.15\right] \\
- \\
- \\
- \\
\end{array}$ \\
\hline
\end{tabular}
\end{center}
\end{table}

%
%
%
\section{New results from the \boldmath{$B$} factories}

\subsection{Measurement of \boldmath{\Dz-\Dzb} mixing and search for indirect \CP violation in \boldmath{$\Dz \to K^+K^-$} and \boldmath{$\Dz \to \pi^+\pi^-$} decays}

The \babar~\cite{casarosa:charm2012} and the Belle experiments~\cite{staric:charm2012} have recently presented updated results for the measurements of the mixing parameter
 $y_{\CP}$ and the \CP violation parameters $\Delta Y$
\footnote{Note that this definition for \deltaY has different sign convention with respect to \babar  
results published previously~\cite{Aubert:2007en,Aubert:2009ai}.}
 for \babar and  $A_{\Gamma}$ for Belle.
The definitions of \deltaY and $A_{\Gamma}$ are the following:
\beq
\deltaY = \frac{\Gamma^+ - \bar{\Gamma}^+}{2\Gamma},  \hspace{0.2cm} 
\AGamma = \frac{\bar{\tau}^+ - \tau^+}{\bar{\tau}^+ + \tau^+} \hspace{0.2cm} \textrm{and} \hspace{0.2cm} \deltaY = (1+\yCP) \AGamma, 
\eeqn
where $\tau^+=1/\Gamma^+$ $(\bar{\tau}^+=1/\bar{\Gamma}^+)$ are the effective lifetimes for \Dz (\Dzb) 
decaying to the \CP-even final states $K^+K^-$ and $\pi^+\pi^-$. 
In principle the parameters \yCP and \deltaY depend on the final state $f$, as indicated in Eq.~(\ref{eq:ycp_f}) 
 and Eq.~(\ref{eq:Agamma_f}) below, when accounting for direct \CP violation.
The \babar and Belle analyses assume \CP conservation in the decay (\ie\ $A_\pi=0$, $A_K=0$),
 and neglect terms of ${\cal O}(10^{-4})$  in the expressions below, that are beyond the present experimental sensitivity:
\beqa
\label{eq:ycp_f}
&\yCP^{f}& = y \cos \phi_f +\frac12 (A_M+A_f ) x\sin \phi_f - \frac14 A_M A_f  y \cos \phi_f  \\ \nonumber 
\Rightarrow &\yCP& \simeq y \cos\phi + \frac12 A_M x\sin \phi \\ 
\label{eq:Agamma_f}
&\deltaY^{f} & = -x  \sin \phi_f +\frac12 (A_M+A_f ) y\cos \phi_f + \frac14 A_M A_f x \sin \phi_f \\ \nonumber
\Rightarrow &\deltaY &\simeq -x  \sin \phi +\frac12 A_M y\cos \phi.
\eeqan
Hence, the parameters no longer depend on the final state $f$, and so can be averaged over the $K^+K^-$ and $\pi^+\pi^-$ modes.

 The measurements are based on the ratio of lifetimes extracted simultaneously from a sample of \Dz mesons produced
 through the flavor-tagged  process $D^{*+} \to \Dz \pi^+$, where the \Dz decays to $K^-\pi^+$, $K^-K^+$, 
or $\pi^-\pi^+$;
 \babar uses the additional samples of untagged  decays $\Dz \to K^-\pi^+$ and $\Dz \to K^-K^+$ for the measurement 
 of \yCP. These have about 4 times the  statistics of the corresponding flavor-tagged sample, but have 
lower purity.
 
The main selection criteria require that the center-of-mass momentum of the \Dz ($p^*$) be 
greater that 2.5 \gevc
\footnote{Belle requires $p^*>3.1~\gevc$  for the data collected at the $\Upsilon(5S)$ mass peak.}, 
 that \Dz daughter tracks be identified as kaons and pions,
 and require for the flavor-tagged sample  that the reconstructed $\Deltam=m(D^{*+})-m(D^0)$ 
be close to the value of 0.1455 \gev.
 The selection criterion on $p^*$ is used to reject $D$ mesons from $B$ decays, and to improve 
the signal significance.
 
The proper time $t$ and proper-time error $\sigma_t$ are 
obtained from the reconstruction of the 3-dimensional 
flight length $( \vec{L})$ and the momentum of the \Dz ($\vec{p}$) according to the relation $t = m/|\vec{p}|^2 \vec{L} \cdot \vec{p} $, 
where $m$ is the nominal mass of the \Dz.
The flight length is reconstructed by means of a kinematic fit to the decay vertex and production vertex 
of the \Dz,
 the latter being constrained 
to  originate within the \epem collision region.
The typical transverse dimensions of the luminous region of the PEP-II collider are about 100 $\mu m$ in the \x direction and 
7 $\mum$ in the \y direction. The most probable $\sigma_t$ value is about 40\% of the nominal \Dz lifetime, and
only candidates with 
 $\sigma_t<0.5 \ps$ are retained for the fit in the \babar analysis.

 The lifetimes of the \CP-even modes $K^-K^+,~\pi^-\pi^+$ are compared to that of the \CP-mixed mode $K^-\pi^+$ in order to measure 
 $y_{\CP}$, which is proportional to the ratio of the lifetimes, and $\Delta Y$ ($A_\Gamma$) which is proportional to the difference 
of the effective lifetimes of \Dzb and \Dz into \CP-even modes.
 \babar measures  $y_{\CP} = [0.72 \pm 0.18(\textrm{stat}) \pm 0.12(\textrm{syst}) ] \%$ and 
$\Delta Y = [0.09 \pm 0.26(\textrm{stat}) \pm 0.06(\textrm{syst}) ] \%$ using a data sample corresponding 
 to an integrated luminosity of 468 $fb^{-1}$~\cite{casarosa:charm2012}.
Belle measures 
$y_{\CP} = [1.11 \pm 0.22(\textrm{stat}) \pm 0.11(\textrm{syst}) ] \%$ and 
$A_{\Gamma} = [-0.03 \pm 0.20(\textrm{stat}) \pm 0.08(\textrm{syst}) ] \%$
using a data sample corresponding to 976 $fb^{-1}$~\cite{staric:charm2012}. 

The systematic uncertainties on \yCP and \deltaY  are reported in Table~\ref{tab:syst_babar} for \babar, and 
 on \yCP and \AGamma are reported in Table~\ref{tab:syst_belle} for Belle. 
%
%
\begin{table}[t]
\begin{center}
\begin{tabular}{|l|cc|}  
\hline
Category  & $\Delta \yCP~(\%)$  & $\Delta(\deltaY)~(\%)$ \\
\hline
Fit region        &  0.057 & 0.022  \\
Signal model      &  0.022 & 0.000  \\
Charm bkg         &  0.045 & 0.001  \\
Combinatorial bkg &  0.079 & 0.002  \\
Selection         &  0.059 & 0.054  \\
\hline
Total             &  0.124 & 0.058  \\
\hline
\end{tabular}
\caption{Systematic uncertainties for \yCP and \deltaY at \babar. The
total is the sum-in-quadrature of the entries in each column.}
\label{tab:syst_babar}
\end{center}
\end{table}
%
\begin{table}[t]
\begin{center}
\begin{tabular}{|l|cc|}  
\hline
Category  & $\Delta \yCP~(\%)$  & $\Delta \AGamma~(\%)$ \\
\hline
Acceptance           &  0.050 & 0.044  \\
SVD misalignments    &  0.060 & 0.041  \\
Mass window position &  0.007 & 0.009  \\
Background           &  0.059 & 0.050  \\
Resolution function  &  0.030 & 0.002  \\
\hline
Total                &  0.11 & 0.08  \\
\hline
\end{tabular}
\caption{Systematic uncertainties for \yCP and \AGamma at Belle. The
total is the sum-in-quadrature of the entries in each column.}
\label{tab:syst_belle}
\end{center}
\end{table}
%
%
The total systematic uncertainties are comparable between the two experiments. It is worth noting that 
the Belle experiment quotes a systematic error due to the silicon vertex detector (SVD) misalignment that is 
  is negligible in the case of the \babar experiment.
The measurement of the proper time average value in Belle shows a dependence on the cosine of the polar angle 
in the \epem center-of-mass ($\cos \theta^*$) that is not properly reproduced by Monte Carlo simulation. 
The discrepancy between data and Monte Carlo is different for the data collected with the two vertex detectors
 that have been used during the running of the experiment. 
In order to minimize the systematic error due to this effect, Belle has performed the 
 measurement by dividing the sample into different intervals of $\cos \theta^*$. 

The $y_{\CP}$ measurements report evidence for \Dz-\Dzb mixing with a significance of 
  $3.3\sigma$ in the case of \babar (most precise measurement to date) and $4.5\sigma$ in the case of Belle. 
The measurements of $\deltaY$ from \babar, and $\AGamma$ from Belle are compatible within error and both are 
consistent with no \CPV.
The new measurements supersede the previous \babar~\cite{Aubert:2007en,Aubert:2009ai} 
and Belle results~\cite{Staric:2007dt}.
The updated HFAG averages, including the new results, are $\yCP = (0.866 \pm 0.155)\%$ and $\AGamma = (-0.022 \pm 0.161 ) \%$.
The comparison  with the previous HFAG average values $\yCP = (1.064 \pm 0.209)\%$ and $\AGamma = (0.026 \pm 0.231 ) \%$ indicates significant improvement in  precision and a lower central value for \yCP.

\section{Summary}
 
In summary, HFAG provides averages for the \Dz-\Dzb mixing and \CP-violating parameters by combining results 
from the following experiments: \babar, Belle, CDF, CLEO, CLEOc, E791, FOCUS, and LHCb.
The average values from the \CPV-allowed fit for the mixing parametes are $x = (0.63^{+ 0.19}_{- 0.20})\%$, $y = (0.72 \pm 0.12 )\%$,  
and for the \CP-violating parameters are $\absqop = 0.88^{+ 0.18}_{- 0.16}$, $\phi= (-10.1^{+ 9.5}_{- 8.9})^\circ$.
These average values do not include the new \babar and Belle results presented here.
The parameter \x differs from zero by $2.7\sigma$, and \y differs from zero by $6.0\sigma$.
The no-mixing hypothesis, $(x=0, y=0)$, is excluded with a statistical significance of $10.2\sigma$.
The LHCb and CDF experiments have obtained first evidence for direct \CPV in \Dz
decays, while there is no evidence for \CPV in mixing $(\absqop \ne 1)$ and in the interference between mixing and decay $(\phi \ne 0)$.
 
New results for \Dz-\Dzb mixing and \CPV in the lifetime ratio analysis of the transitions \dkkpp\ to the 
 transitions $\Dz \to K^-\pi^+$ have been presented.
\babar measures 
$y_{\CP} = [0.72 \pm 0.18(\textrm{stat}) \pm 0.12(\textrm{syst}) ] \%$ and 
$\Delta Y = [0.09 \pm 0.26(\textrm{stat}) \pm 0.06(\textrm{syst}) ] \%$ using a data sample 
 corresponding to an integrated luminosity of 468 $fb^{-1}$.
Belle measures 
$y_{\CP} = [1.11 \pm 0.22(\textrm{stat}) \pm 0.11(\textrm{syst}) ] \%$ and 
$A_{\Gamma} = [-0.03 \pm 0.20(\textrm{stat}) \pm 0.08(\textrm{syst}) ] \%$
using a data sample corresponding to 976 $fb^{-1}$. 
The updated HFAG averages, including the new results, are $\yCP = (0.866 \pm 0.155)\%$ and $\AGamma = (-0.022 \pm 0.161 ) \%$.

HFAG averages for \Dz-\Dzb mixing and \CPV parameters are in agreement with the SM. 
Recent results for direct \CPV in \dkkpp\ decays report a larger asymmetry value than the SM expectation.
However, given the present knowledge of the
charm system, this cannot (yet) be considered a clear signal of physics beyond the SM.

\Acknowledgements
I am grateful to my \babar collegues for providing new results in time for the workshop and for useful comments.
I would like to thank the conference organizers for their warm hospitality in a wonderful location.

\end{document}